\documentclass{mn2e}

\usepackage{amsfonts}
\usepackage{amsmath}
\usepackage{graphicx}

\title[Plain fundamentals of Fundamental Planes]
      {Plain fundamentals of Fundamental Planes:  Analytics and algorithms}

\author[R. K. Sheth \& M. Bernardi]
{Ravi K. Sheth$^{1,2}$ \& Mariangela Bernardi$^2$\thanks{E-mail: shethrk,bernardm@physics.upenn.edu}\\
 $^1$ The Abdus Salam International Center for Theoretical Physics, 
      Strada Costiera 11, 34151 Trieste, Italy\\
 $^2$ Department of Physics \& Astronomy, University of Pennsylvania, 
      209 S. 33rd St., Philadelphia, PA 19104, USA}

\newcommand{\bm}[1]{{\mbox{\boldmath $#1$}}}

\begin{document}
\pagerange{\pageref{firstpage}--\pageref{lastpage}}

\maketitle

\label{firstpage}

\begin{abstract}
Estimates of the coefficients $a$ and $b$ of the Fundamental Plane 
relation $R\propto \sigma^a\,I^b$ depend on whether one minimizes 
the scatter in the $R$ direction, or orthogonal to the Plane.  
We provide explicit expressions for $a$ and $b$ 
(and confidence limits) in terms of the covariances between 
$\log R$, $\log \sigma$ and $\log I$.  
Our expressions quantify the origin of the difference between 
the direct, inverse and orthogonal fit coefficients.  
They also show how to account for correlated errors, 
how to quantify the difference between the Plane in a magnitude 
limited survey and one which is volume limited, how to determine 
whether a scaling relation will be biased when using an apparent 
magnitude limited survey, how to remove this bias, and why some forms 
of the $z\approx 0$ Plane appear to be less affected by selection effects, 
but that this does not imply that they will remain unaffected at high 
redshift.  
Finally, they show why, to a good approximation, the three vectors 
associated with the Plane, one orthogonal to and the other two in it, 
can all be written as simple combinations of $a$ and $b$.  
Essentially, this is a consequence of the fact that the distribution 
of surface brightnesses is much broader than that of velocity dispersions, 
and velocity dispersion and surface brightness are only weakly correlated.  
Why this should be so for galaxies is a fundamental open question 
about the physics of early-type galaxy formation.  
We argue that, if luminosity evolution is differential, and sizes 
and velocity dispersions do not evolve, then this is just an accident: 
velocity dispersion and surface brightness must have been correlated in 
the past.  On the other hand, if the (lack of) correlation is similar 
to that at the present time, then differential luminosity evolution 
must have been accompanied by structural evolution.  
A model in which the luminosities of low luminosity galaxies 
evolve more rapidly than do those of higher luminosity galaxies is able 
to produce the observed decrease in $a$ (by a factor of 2 at $z\sim 1$) 
while having $b$ decrease by only about 20 percent.  In such a model, 
the dynamical mass-to-light ratio is a steeper function of mass at 
higher $z$.  
Our analysis is more generally applicable to any other correlations 
between three variables:  e.g., the color-magnitude-$\sigma$ relation, 
the luminosity and velocity dispersion of a galaxy and the mass of its 
black-hole, or the relation between the X-ray luminosity, Sunyaev-Zeldovich 
decrement and optical richness of a cluster, so we provide {\tt IDL} 
code which implements these ideas.  
And, for completeness, we show how our analysis generalizes further 
to correlations between more than three variables.  
\end{abstract}

\begin{keywords}
methods: analytical - methods: statistical - 
galaxies: formation - galaxies: fundamental parameters
\end{keywords}

\section{Introduction}
Early-type galaxies do not fill the full three dimensional space 
defined by size, central velocity dispersion and surface brightness 
(usually evaluated at the half light radius).  Rather, they define 
a relatively thin manifold which has come to be called the 
Fundamental Plane (e.g. Djorgovski \& Davis 1987; 
J{\o}rgensen et al. 1996; Pahre et al. 1998; Bernardi et al. 2003; 
J{\o}rgensen et al. 2006; Bolton et al. 2008; Hyde \& Bernardi 2009b). 

The Fundamental Plane is usually written as 
\begin{equation}
 \log_{10}\frac{R_e}{{\rm kpc}} = a\,\log_{10}\frac{\sigma}{{\rm km~s^{-1}}} 
                          - \frac{b}{2.5}\,\frac{\mu_e}{{\rm mags}} + c,
\end{equation}
where $R_e$ is the half light radius, $\sigma$ is the velocity 
dispersion (typically corrected to an aperture of $R_e/8$), 
and $\mu_e$ is the surface brightness within $R_e$.  
The coefficient $a$ is loosely refered to as the `slope', 
and $c$ is the `zero-point'; it is simply 
$c = \langle \log_{10}R\rangle - a\,\langle\log_{10}\sigma\rangle 
                               + 0.4b\,\langle\mu_e\rangle$.
The shape of the Fundamental Plane is determined by estimating $a$ 
and $b$.  
The values of $a$ and $b$ are thought to encode useful information 
about these objects.  This is because the values $a=2$ and $b=-1$ are 
expected on dimensional grounds if the virial theorem holds exactly 
in the observed variables, and mass is linearly proportional to light.  

If $a\ne 2$ and/or $b\ne -1$ then the FP is said to be `tilted'. 
The tilt may be due to a combination of stellar population effects, 
initial mass function variations, and variations in the dark matter 
fraction within $R_e$ (e.g. Pahre et al. 1998; Bernardi et al. 2003; 
Bolton et al. 2008; Hyde \& Bernardi 2009b; Graves \& Faber 2010). 
However, the inferred tilt also depends on how the parameters $a$ and $b$ 
were measured.  This is typically done either by minimizing residuals in 
the $R_e$ direction, or in the direction orthogonal to the fit.  
In general the `direct' and `orthogonal' fit parameters are 
different combinations of the mean values of and covariances 
between the variables $\log_{10}R$, $\log_{10}\sigma$ and $\mu$.  
Moreover, in practice, naive estimation of these means and 
covariances (e.g. simply summing over the data without including 
other weight terms) may lead to biases induced by measurement errors 
(these usually affect the covariances) or by selection effects 
(which bias the means and the covariances).  
The effects of both must be accounted-for to estimate the 
intrinsic shape parameters $a$ and $b$ (e.g. Saglia et al. 2001).  
This is especially important when the FP is determined for galaxies 
in a magnitude limited sample (Bernardi et al. 2003).  

The main goal of this paper is to provide analytic expressions which 
describe the Plane for both the direct, inverse and orthogonal fitting 
procedures which show clearly how to account for measurement errors 
and selection effects. In addition, by providing analytic expressions 
for all quantities of interest, our results remove the need for
numerical nonlinear minimization methods for obtaining the best-fit 
coefficients. Our analysis is complementary to that in 
Saglia et al. (2001), who provide an excellent description of the 
key differences between the different fitting procedures.  When we 
illustrate the results of our analysis, the numerical values we use 
come from the SDSS-based early-type sample compiled by 
Hyde \& Bernardi (2009b).  

The discussion above has focussed on the direction of the smallest 
scatter.  If we think of the Plane as being defined by three 
orthogonal vectors, one orthogonal to the Plane and the others in 
it, then the parameters $a$ and $b$ describe the vector which is 
orthogonal to the plane.  If ${\bm\Lambda}_3$ denotes this vector, 
and the other two vectors (in the Plane) are ${\bm\Lambda}_1$ and 
${\bm\Lambda}_2$, then Saglia et al. (2001) showed that these three 
eigenvectors are well-approximated by
\begin{eqnarray}
 {\bm\Lambda}_3 &=& {\bm r} - a_{\rm orth}\, {\bm v} - b_{\rm orth}\, {\bm i}
  \nonumber\\
 {\bm\Lambda}_2 &\approx& {\bm r} 
                         + \frac{(1 + b_{\rm orth}^2)}{a_{\rm orth}}\, {\bm v} 
                         - b_{\rm orth}\, {\bm i} \nonumber\\
 {\bm\Lambda}_1 &\approx& {\bm r} + b_{\rm orth}^{-1}\, {\bm i},
 \label{FPvectors}
\end{eqnarray}
where ${\bm r}$, ${\bm v}$, and ${\bm i}$ denote unit vectors in 
the size, velocity dispersion and surface brightness directions.  
Although Saglia et al. justified these scalings using numerical 
experiments, we show, in Section~\ref{analysis}, that this form 
follows from the fact that the distribution of surface brightnesses 
is much broader than that of velocity dispersions.  

Section~\ref{analysis} also shows that many of the properties of the 
$z=0$ Fundamental Plane can be understood as arising from the fact 
that surface brightness and velocity dispersion are almost uncorrelated 
at $z=0$.  In Section~\ref{evolve} we argue that, in models of pure 
luminosity evolution, this is only a coincidence:  the two were 
correlated in the past.  A final section summarizes our conclusions 
and discusses why measurements of this correlation in high-$z$ datasets 
will provided interesting constraints on models.  

In an Appendix, we provide a description of how the FP coefficients 
differ between magnitude limited and volume limited samples, when the 
underlying pairwise scaling relations are linear.  Although there 
is now growing evidence for curvature in these relations 
(e.g. Bernardi et al. 2007a; Lauer et al. 2007; Hyde \& Bernardi 2009a; 
Bernardi et al. 2011), we feel our expressions are useful since the 
curvature is usually due to a small fraction of the objects in the 
tails of the distribution. Moreover, our expressions are generally 
applicable to any study of three observables -- not just those 
associated with the Fundamental Plane.  It may be that the assumption 
of no curvature is more accurate for some of these other scaling 
relations.  
Some examples include the joint distribution of the luminosity 
and velocity dispersion of a galaxy and its color or the mass of its 
black-hole (Bernardi et al. 2005; Bernardi et al. 2007b), or the 
relation between the X-ray luminosity, SZ-signal strength and optical 
richness of a cluster.

\section{Analytic description of the Fundamental Plane}\label{analysis}
The analysis which follows is actually the restriction to a special 
case of the following general statement.  Since the general case is 
also of interest in these glorious days of large panchromatic datasets, 
we state it first.

\subsection{Conditional correlations between $N$ variables}
Suppose we have $N$ observables which are distributed following a 
multivariate Gaussian distribution having means $\mu_i$ and covariance 
matrix $C_N$.  Suppose that we split them up into two sets, A with $n$ 
observables and B with the other $N-n$.  Let $\mu_A$ and $C_{AA}$ denote 
the mean vector and covariance matrix of set A, and similarly define 
$\mu_B$ and $C_{BB}$ for set B.  Then the distribution of
 $O_A = \{X_1,\ldots,X_n\}$ given that $O_B = \{X_{n+1},\ldots,X_N\}$ 
is known, is multivariate Gaussian with mean 
\begin{equation}
 \mu_{A|B} = \langle O_A|O_B\rangle = \mu_A + C_{AB}C_{BB}^{-1}\,(O_B-\mu_B),
 \label{condMean}
\end{equation}
and covariance matrix 
\begin{equation}
 C_{A|B} = C_{AA} - C_{AB}\,C^{-1}_{BB}\,C_{BA}.
 \label{condCovariance}
\end{equation}
In what follows, we will study the special case in which $N=3$ and $n=1$.  
Since this makes $C_{BB}$ a $2\times 2$ matrix, its inverse is simple, 
so the expression above is analytically tractable.

\subsection{Restriction to $N=3$}
For our three variables, we will use $R$, $V$ and $I$ to denote 
$\log (R$/kpc), $\log (\sigma$/km~s$^{-1}$) and 
$\log (I/(L_\odot{\rm pc}^{-2}))$.  
Let ${\cal C}$ denote the real symmetric matrix which describes 
the covariances between these three variables:
\begin{equation}
 {\cal C} \equiv \left(\begin{array}{ccc}
           C_{II} & C_{IR} & C_{IV}\\
           C_{IR} & C_{RR} & C_{RV}\\
           C_{IV} & C_{RV} & C_{VV}\\
          \end{array}\right).
 \label{calF}
\end{equation}
The shape of the Fundamental Plane is completely determined by 
this covariance matrix.  Hence, our problem is to estimate the 
coefficients of this matrix in a way which accounts for selection 
effects and measurement errors (see Section~\ref{covar}).  

In what follows, we will provide expressions for various quantities 
which can be derived from ${\cal C}$.  Although our expressions are 
general, we will sometimes remark on what they imply.  In such cases, 
we will use the values reported by Hyde \& Bernardi (2009b):
\begin{equation}
{\cal C} = 
 \left(\begin{array}{rrr}
           0.0471 & -0.0313 & 0.0038\\
          -0.0313 &  0.0552 & 0.0189\\
           0.0038 &  0.0189 & 0.0187\\
          \end{array}\right),
 \label{Fhb09}
\end{equation}
where $I$ was measured in dex (rather than magnitudes).  
In particular, Table~\ref{tab:FPfits} summarizes the various values 
of $a$ and $b$ which can be derived from this ${\cal C}$, depending on 
how one fits the Fundamental Plane.  Note that these coefficients are 
often determined via numerical nonlinear minimization schemes.  
In the following subsections, we provide analytic expressions for 
these parameters, thus eliminating the need for such schemes.

\begin{table}
 \caption[]{Coefficients of various fits to the Fundamental Plane 
            $R\propto \sigma^a I^b$ in the $r$-band sample of 
            about 40000 objects defined by Hyde \& Bernardi (2009b), 
            after correcting for the magnitude limit selection effect.  
            Confidence limits ignore the contribution from systematic errors.}
 \centering
 \begin{tabular}{lcc}
  \hline 
    & $a$ & $b$ \\
  \hline 
  Direct  & $1.167\pm 0.014$ & $-0.757\pm 0.009$\\
  Inverse & $1.606\pm 0.023$ & $-0.792\pm 0.010$\\
  SB      & $1.219\pm 0.017$ & $-1.028\pm 0.009$ \\
  Orthogonal & $1.434\pm 0.015$ & $-0.787\pm 0.010$ \\
  \hline \\
\end{tabular}
\label{tab:FPfits} 
\end{table}

Note that $|C_{IV}|$ is the smallest element of ${\cal C}$.  
To remove the effect of the fact that the rms of $I$ is much 
larger than that in $R$ or $V$ (and depends on whether $I$ is 
measured in dex or in mags!), we can normalize all quantities by 
their rms values.  If we define
\begin{equation}
 r_{xy} \equiv \frac{C_{xy}}{\sqrt{C_{xx}C_{yy}}}
\end{equation}
and call the resulting covariance matrix ${\cal R}$, then 
\begin{equation}
{\cal R} = 
 \left(\begin{array}{rrr}
           1  & -0.614 & 0.128\\
          -0.614  & 1 &  0.588\\
           0.128  & 0.588 & 1 \\
          \end{array}\right).
 \label{calR}
\end{equation}
This shows that $r_{IV}$ is indeed much smaller than $r_{IR}$ 
or $r_{RV}$:  surface brightness and velocity dispersion are 
almost uncorrelated.  This turns out to be a simple way to 
understand many features of the Fundamental Plane.  

\subsection{Accounting for selection effects and measurement errors}
\label{covar}
In an apparent magnitude limited survey of $N_{\rm obj}$ objects, 
the mean value of an observed quantity $X$,
 $\bar X \equiv \sum_i^{N_{\rm obj}} X_i/N_{\rm obj}$, 
may be biased from its true mean value (e.g., if the observable 
correlates with luminosity).  Fortunately, this bias is easily 
removed by defining, for each object with luminosity $L_i$, the total 
volume over which the object could have been observed: $V_{\rm max}(L_i)$ 
(e.g. Schmidt 1968).  One then uses this to define a (normalized) weight 
\begin{equation}
 w_i = \frac{V_{\rm max}^{-1}(L_i)}{\sum_i V_{\rm max}^{-1}(L_i)},
\end{equation}
and estimates the mean value of $X$ as 
\begin{equation}
 \langle X\rangle = \sum_i w_i\,X_i,
 \label{meanX}
\end{equation}
where the sum is over all the objects in the sample.  

For similar reasons, the covariance between observables will also 
be biased by the selection effect, but this bias can be removed by 
applying the same weight.  The covariance may also be biased by 
measurement errors.  If we define the matrix $\cal{O}$ to have 
elements 
\begin{equation}
 O_{XY} = \sum_i w_i\left[\Bigl(X_i - \langle X\rangle\Bigr)\,
                      \Bigl(Y_i - \langle Y\rangle\Bigr)\right],
\end{equation}
and the measurement error matrix $\cal{E}$ by 
\begin{equation}
 E_{XY} = \sum_i w_i\, \langle e_Xe_Y\rangle_i
 \label{Exy}
\end{equation}
(we have assumed zero mean for the errors, and often, 
$\langle e_Xe_Y\rangle_i$ is assumed to be the same for all objects),
then
\begin{equation}
 \cal{C} = \cal{O} - \cal{E}
 \label{Cxy}
\end{equation}
is an unbiased estimate of the intrinsic covariance matrix.
Notice that each element of $\cal{C}$ has had the contribution from 
measurement errors to the observed covariance subtracted off:  
 $C_{XY} = O_{XY} - E_{XY}$.  
If this term is not subtracted, i.e., if one uses ${\cal O}$ instead 
of ${\cal C}$ in what follows, one will obtain a Plane that has been 
distorted by measurement error.  
In the Appendix, we quantify the bias which results from ignoring 
the $V_{\rm max}^{-1}$ weight; i.e., of setting $w=1/N_{\rm obj}$ for 
all $i$.

Some workers like to account for the fact that certain measurements 
are more secure than others by weighting each measurement by the 
inverse of the estimated uncertainty on it.  In this case, if one 
defines 
\begin{equation}
 O^{\rm E}_{XY} = 
         \frac{\sum_i w_i\,
                 \frac{(X_i - \langle X\rangle)}{\sqrt{\langle e_X^2\rangle_i}}
                 \frac{(Y_i - \langle Y\rangle)}{\sqrt{\langle e_Y^2\rangle_i}}}
              {\sum_i w_i/\sqrt{\langle e_X^2\rangle_i \langle e_Y^2\rangle_i}},
 \label{ewtdOxy}
\end{equation}
where 
\begin{equation}
 \langle X\rangle = \frac{\sum_i w_i\, X_i/\sqrt{\langle e_X^2\rangle_i}}
                         {\sum_i w_i/\sqrt{\langle e_X^2\rangle_i}},
\end{equation}
then one must also define 
\begin{equation}
 E^{\rm E}_{XY} = \frac{\sum_i w_i \langle e_X e_Y\rangle_i
                      /\sqrt{\langle e_X^2\rangle_i \langle e_Y^2\rangle_i}}
           {\sum_i w_i/\sqrt{\langle e_X^2\rangle_i \langle e_Y^2\rangle_i}}
\end{equation}
before estimating 
\begin{equation}
 C_{XY} = O^{\rm E}_{XY} - E^{\rm E}_{XY}
\end{equation}
as before.  In practice, it makes sense to replace 
 $1/\sqrt{\langle e_X^2\rangle}\to 
  1/\sqrt{\epsilon_{\rm min}^2 + \langle e_X^2\rangle}$ 
for some $\epsilon_{\rm min}^2$ that is chosen to prevent a few 
well-measured objects from dominating the sums.  

\subsection{The parameters of the direct fit}
If we write the Fundamental Plane as 
\begin{equation}
 R - \langle R\rangle = a\, \Bigl(V-\langle V\rangle\Bigr) 
                      + b\, \Bigl(I - \langle I\rangle\Bigr),
\end{equation}
then 
\begin{eqnarray}
 \label{adirect}
 a_{\rm direct} &=& \frac{(C_{RV}/C_{VV}) - (C_{IR}/C_{II})(C_{IV}/C_{VV})}
                       {1 - (C_{IV}/C_{II})(C_{IV}/C_{VV})} \\
     &=& \frac{C_{RV}}{C_{VV}} \frac{1 - r_{IV}r_{IR}/r_{RV}}{1 - r_{IV}^2}; \\
 b_{\rm direct} &=& (C_{IR}/C_{II}) -  a_{\rm direct}\,(C_{IV}/C_{II}) \\
         &=& \frac{(C_{IR}/C_{II}) - (C_{IV}/C_{II})(C_{RV}/C_{VV})}
                       {1 - (C_{IV}/C_{II})(C_{IV}/C_{VV})} \\
         &=& \frac{C_{IR}}{C_{II}}\,\frac{1 - r_{IV}r_{RV}/r_{IR}}{1 - r_{IV}^2}
 \label{bdirect}
\end{eqnarray}
(Bernardi et al. 2003).  Note that because of how we defined our $C_{XY}$, 
these expressions have been corrected for the effects of errors, and 
because of the weighting term $w_i$, they have been corrected for 
selection effects.  

Equation~(\ref{adirect}) shows that $a_{\rm direct} $ is simply the 
correlation between $R$ and $V$ minus the contribution which comes from 
$R-I$ and $I-V$ correlations.  Similarly, $b_{\rm direct}$ is the correlation 
between $R$ and $I$ minus the contribution which comes from the $R-V$ 
and $I-V$ correlations.  It might help to think of these as follows.  
Let
 $X_{R|I} \equiv R-\langle R\rangle - (C_{RI}/C_{II})\, (I - \langle I\rangle)$ 
denote the residual in $R$ from the $R-I$ correlation.  Then 
 $\langle X_{R|I}V\rangle = C_{RV} - (C_{RI}/C_{II})\, C_{IV}$.  
Therefore, $a_{\rm direct}$ is the ratio of $\langle X_{R|I}V\rangle$ to 
the range of $V$ values at fixed $I$, $C_{VV}(1-r_{IV}^2)$, so it is 
the slope of the correlation between $X_{R|I}$ and $V$, at fixed $I$.  
Of course, $b_{\rm direct}$ can be understood similarly.  

The fact that, in the data, neither $a_{\rm direct}$ nor $b_{\rm direct}$ 
are zero implies that both the $R-V$ and $I-R$ correlations are 
fundamental -- they are not consequences of other relations.  
Moreover, note that if $C_{IV}=0$ (i.e., $r_{IV}=0$), then $a_{\rm direct}$ 
and $b_{\rm direct}$ are really just the slopes of the 
$\langle R|V\rangle$ and $\langle R|I\rangle$ relations.  
In addition, if $C_{IV}\approx 0$, then the Direct fit has the convenient 
property that the errors on the fitted coefficients $a_{\rm direct}$ and 
$b_{\rm direct}$ are independent.  We show below that $C_{IV}\approx 0$ 
turns out to be an easy way to understand some properties of the 
Fundamental Plane.  

This form of the Plane (i.e., the Direct fit) should be used if the 
distance independent quantities $V$ and $I$ are used to predict the 
distant dependent one $R$.  The accuracy with which $R$ is predicted 
by $I$ and $V$ is limited by the rms scatter around this fit, which is 
(the square root of) 
\begin{equation}
 \langle\Delta R_{\rm direct}^2\rangle
   = C_{RR} 
   \frac{1 - r_{RV}^2 - r_{IV}^2 - r_{IR}^2 + 2 r_{IR}r_{IV}r_{RV}}{1 - r_{IV}^2}.
 \label{chi2min}
\end{equation}

Confidence limits on $a_{\rm direct}$ and $b_{\rm direct}$ themselves can
be obtained as follows.  
If there were no measurement errors, then the 
68\% confidence limits on the best fit values 
$a_{\rm direct}$ and $b_{\rm direct}$ would be given by the square root of 
 $\langle\Delta R_{\rm direct}^2\rangle/[N_{\rm obj}C_{V|I}]$
and 
 $\langle\Delta R_{\rm direct}^2\rangle/[N_{\rm obj}C_{I|V}]$,
where we have defined $C_{Y|X} = C_{YY}(1 - r_{XY}^2)$ and 
 $\langle\Delta R_{\rm direct}^2\rangle$ is given by equation~(\ref{chi2min}). 
Note that the confidence limit on $a_{\rm direct}$ is proportional to 
the scatter around the best fit, $\langle\Delta R_{\rm direct}^2\rangle$, 
divided by the number of degrees of freedom (which is essentially the 
sample size), as one might expect.  However, it is also scales inversely 
with $C_{V|I}$ because, as the intrinsic spread in $V$ at fixed $I$ 
decreases, it becomes increasingly difficult to measure the slope of 
the $R-V$ relation (at fixed $I$).  
Similar arguments apply to $b_{\rm direct}$.
This means that the uncertainty on $a_{\rm direct}$ will be 
$\sqrt{C_{II}/C_{VV}}$ times the uncertainty on $b_{\rm direct}$, 
independent of sample size.  The errors on these best-fitting 
coefficients are correlated.  The correlation is the square root of 
$\langle\Delta R_{\rm direct}^2\rangle\,C_{IV}/[N_{\rm obj}C_{V|I}C_{I|V}]$;
it is nonzero if $C_{IV}\ne 0$.

Measurement errors (random, not systematic) decrease the precision of 
these estimates as follows.  If  
 $\chi^2_{\rm obs,dir} \equiv O_{RR} + a^2_{\rm direct}O_{VV} + b_{\rm direct}O_{II}
                      - 2 a_{\rm direct} O_{RV} - 2 b_{\rm direct}O_{IR} 
                      + 2 a_{\rm direct} b_{\rm direct} O_{IV}$ 
(note that this is just the observational analogue of equation~\ref{chi2min}), 
then the limits on $a_{\rm direct}$ and $b_{\rm direct}$ are well-approximated by 
 $\chi^2_{\rm obs,dir}\, (O^w_{V|I}/C_{V|I})/C_{V|I}$
and
 $\chi^2_{\rm obs,dir}\, (O^w_{I|V}/C_{I|V})/C_{I|V}$,
respectively, where 
 $O^{w}_{Y|Z}\equiv O^w_{YY} - 2(C_{YZ}/C_{ZZ})O^w_{YZ} + (C_{YZ}/C_{ZZ})^2 O^w_{ZZ}$ 
where
 $O^w_{YZ}\equiv \sum_i (w^2 X_{YZ}^2)_i$, 
for $(Y,Z)=(V,I)$ or $(I,V)$ respectively. 

The superscript $w$ is to remind us that $O^w_{Y|Z}$ carries an extra weighting 
factor compared to $O_{Y|Z}$.  
It may be helpful to think of $(O_{Y|Z}/O^w_{Y|Z})$ as defining an effective 
sample size $N_{Y|Z}$.  This is because, if all the weights are the same 
then (because our weights are normalized) $w=1/N_{\rm obj}$, so 
$(O_{Y|Z}/O^w_{Y|Z}) = N_{\rm obj}$.  Thus, the factor $(O^w_{Y|Z}/C_{Y|Z})$ is 
really $(O_{Y|Z}/C_{Y|Z})/N_{Y|Z}$, making the correspondence with the case 
in which there were no measurement errors obvious:  one replaces 
 $\langle\Delta R^2\rangle \to \chi^2_{\rm obs,dir}$ and 
 $C_{Y|Z}\to (C_{Y|Z}/O_{Y|Z})\,C_{Y|Z}$ (to account for measurement errors) 
and 
 $N\to N_{Y|Z}$ (to account for the weights).  If each measurement was 
weighted by its uncertainty, then all $O_{XY}\to O^{\rm E}_{XY}$, 
and all $O^w_{XY}$ are given by equation~(\ref{ewtdOxy}) with $w_i^2$ 
in the sum in the numerator, but only $w_i$ in the denominator. 

\subsection{The parameters of the inverse fit}
Some authors prefer to keep the spectroscopic quantity $V$ as the 
dependent variable, and so fit 
\begin{equation}
 V - \langle V\rangle = \frac{R-\langle R\rangle}{a_{\rm inv}}
           - \frac{b_{\rm inv}}{a_{\rm inv}}\, \Bigl(I - \langle I\rangle\Bigr).
 \label{abinverse}
\end{equation}
This has some merit, because the measurement of $V$ is often much 
noiser than that of the combination of $R$ and $I$ which defines 
the Plane (e.g. correlated errors in $R$ and $I$ when fitting to 
the surface brightness profile mean that $0.3\mu - R$ is typically 
determined to within 0.005).  If the errors are essentially all on 
$V$, then they do not bias the coefficients of the `direct' fit to 
this relation, so one can safely ignore them when estimating the 
coefficients of the fit.  

So, the question arises as to how well $(a_{\rm inv},b_{\rm inv})$ 
approximate $(a_{\rm direct},b_{\rm direct})$.  By simply interchanging 
$R$ and $V$ in the expressions above, one finds 
\begin{eqnarray}
 \label{ainverse}
 a_{\rm inv} &=& \frac{1 - (C_{IR}/C_{II})(C_{IR}/C_{RR})}
                    {(C_{RV}/C_{RR}) - (C_{IV}/C_{II})(C_{IR}/C_{RR})} \\
    &=& \frac{C_{RR}}{C_{RV}}\frac{1 - r_{IR}^2}{1 - r_{IV}r_{IR}/r_{RV}} \\
       &=& a_{\rm direct}\, \frac{(1 - r_{IR}^2)(1-r_{IV}^2)}
                    {(r_{RV} - r_{IV}r_{IR})^2},\\
 b_{\rm inv} &=& -a_{\rm inv}\,\frac{(C_{IV}/C_{II}) - (C_{IR}/C_{II})(C_{RV}/C_{RR})}
                               {1 - (C_{IR}/C_{II})(C_{IR}/C_{RR})} \\
           &=& -\frac{(C_{IV}/C_{II}) - (C_{IR}/C_{II})(C_{RV}/C_{RR})}
                    {(C_{RV}/C_{RR}) - (C_{IV}/C_{II})(C_{IR}/C_{RR})} \\
           &=& b_{\rm direct}\frac{(r_{IR}r_{RV} - r_{IV})(1-r_{IV}^2)}
                    {(r_{RV} - r_{IV}r_{IR})(r_{IR}-r_{IV}r_{RV})},
 \label{binverse}
\end{eqnarray}
with rms scatter equal to the square root of 
\begin{equation}
 \langle\Delta V_{\rm inv}^2\rangle = C_{VV} \,
   \frac{1 - r_{RV}^2 - r_{IV}^2 - r_{IR}^2 + 2 r_{IR}r_{IV}r_{RV}}{1 - r_{IR}^2}.
\end{equation}
The intrinsic uncertainty on $(1/a_{\rm inv})$ is 
 $\langle\Delta V_{\rm inv}^2\rangle^{1/2}/[N_{\rm obj}C_{R|I}]^{-1/2}$, 
and that for $(b_{\rm inv}/a_{\rm inv})$ is $(C_{RR}/C_{II})^{1/2}$ times that 
on $1/a_{\rm inv}$.
However, the uncertainties on $a_{\rm inv}$ and $b_{\rm inv}$ themselves are 
$\langle\Delta V_{\rm inv}^2\rangle^{1/2}\,a_{\rm inv}^2/[N_{\rm obj}C_{R-b_{\rm inv}I}]^{-1/2}$ 
where 
 $C_{R-b_{\rm inv}I} \equiv C_{RR} - 2b_{\rm inv}C_{IR} + b_{\rm inv}^2 C_{II}$,
and 
 $\langle\Delta V_{\rm inv}^2\rangle^{1/2}\,a_{\rm inv}/[N_{\rm obj}C_{II}]^{-1/2}$.
As before, a good estimate of the uncertainties in the presence of measurement errors and weights comes from replacing 
 $\langle\Delta V_{\rm inv}^2\rangle\to \chi^2_{\rm obs,inv}$, 
 $C_{R-b_{\rm inv}I}\to C_{R-b_{\rm inv}I}(C_{R-b_{\rm inv}I}/O_{R-b_{\rm inv}I})$ 
 and $N_{\rm obj}\to O_{R-b_{\rm inv}I}/O^w_{R-b_{\rm inv}I}$ for $a_{\rm inv}$ and 
 $N_{\rm obj}C_{II}\to C_{II}\,(C_{II}/O^w_{II})$ for $b_{\rm inv}$.

Notice that, in general,
 $a_{\rm inv}\ne a_{\rm direct}$ and $b_{\rm inv}\ne b_{\rm direct}$.  
E.g., if $C_{IV}\to 0$ then 
\begin{equation}
 a_{\rm inv}\to a_{\rm direct}\,\frac{1 - r_{IR}^2}{r_{RV}^2} 
 \qquad {\rm and}\qquad b_{\rm inv} \to b_{\rm direct}.
 \label{abinv}
\end{equation}
The determinant of ${\cal C}$ (the matrix defined in equation~\ref{calF}) 
must be positive definite, so if $C_{IV}=0$, then 
 $1 - r_{IR}^2 - r_{RV}^2 \ge 0$, 
which means 
 $|a_{\rm inv}| \ge |a_{\rm direct}|$.  
Thus, although $b_{\rm inv} = b_{\rm direct}$ in this limit, 
$a_{\rm inv} \ne a_{\rm direct}$.  Therefore, the temptation to 
rearrange equation~(\ref{abinverse}) so as to use 
$a_{\rm inv}V + b_{\rm inv}I$ to estimate $R$ should be avoided, as 
it is guaranteed to lead to a bias.  
In addition to a bias, the associated noise in this estimator of $R$,
\begin{eqnarray}
 \langle\Delta R_{\rm inv}^2\rangle &=& 
    C_{RR} + a_{\rm inv}^2C_{VV} + b_{\rm inv}^2C_{II} - 2 a_{\rm inv}C_{RV}\nonumber\\
 &&       \qquad - 2b_{\rm inv}C_{IR} + 2 a_{\rm inv}b_{\rm inv}C_{IV},
\end{eqnarray}
is larger than $\langle \Delta R_{\rm direct}^2\rangle$.

\subsection{The SB fit:  Predicting $I$ from $R$ and $V$}
For completeness (though see Graves \& Faber 2010 for why this might 
be an interesting choice), we now give the result of fitting the Plane 
when $I$ is the dependent variable:
\begin{equation}
 I - \langle I\rangle = \frac{R-\langle R\rangle}{b_{\rm I}}
           - \frac{a_{\rm I}}{b_{\rm I}}\, \Bigl(V - \langle V\rangle\Bigr).
\end{equation}
In this case, 
\begin{eqnarray}
 b_{\rm I} &=& \frac{1 - (C_{RV}/C_{VV})(C_{RV}/C_{RR})}
                    {(C_{IR}/C_{RR}) - (C_{IV}/C_{VV})(C_{RV}/C_{RR})} \\
 a_{\rm I} 
         &=& \frac{C_{RV}}{C_{VV}}\,\frac{(C_{IR}/C_{RR}) - (C_{IV}/C_{RV})}
                              {(C_{IR}/C_{RR}) - (C_{IV}/C_{VV})(C_{RV}/C_{RR})} ,
 \label{abI}
\end{eqnarray}
the intrinsic error on $a_{\rm I}$ is $\langle\Delta I_{\rm I}^2\rangle^{1/2}\, b_{\rm I}/[N_{\rm obs}C_{VV}]^{-1/2}$
and on $b_{\rm I}$ is
 $\langle\Delta I_{\rm I}^2\rangle^{1/2}\,b_{\rm I}^2[N_{\rm obj}C_{R-a_{\rm I}V}]^{-1/2}$, with the usual replacements to account for measurement errors.  


It is straightforward to verify that, like the inverse fit, 
 $a_{\rm I}(V-\langle V\rangle) + b_{\rm I}(I-\langle I\rangle)$ 
is also a biased predictor of $R-\langle R\rangle$.  
E.g., if $r_{IV}=0$, then $a_{\rm I} \to a_{\rm direct}$ but 
$b_{\rm I}\to b_{\rm direct} \, (1 - r_{RV}^2)/r_{IR}^2$ so 
$|b_{\rm I}| \ge |b_{\rm direct}|$.

\subsection{The orthogonal fit: Eigenvalues}
The expression for the orthogonal fit coefficients is more complicated, 
since it requires knowledge of the eigenvalues and eigenvectors of the 
matrix ${\cal C}$.  However, the eigenvalues of a matrix are the roots 
of its characteristic polynomial, and, since ${\cal C}$ is a 
$3\times 3$ matrix, this polynomial is a cubic, so the roots satisfy
\begin{equation}
 -\lambda^3 + \lambda^2 {\rm Tr}\,{\cal C} 
            + \frac{\lambda}{2}\, [{\rm Tr}\,{\cal C}^2 - {\rm Tr}^2{\cal C}]
            + {\rm Det}\,{\cal C} = 0.
\end{equation}
This can be solved analytically:
since ${\cal C}$ is real and symmetric, the roots are 
\begin{eqnarray}
 \lambda_1 &=& -2\,\sqrt{Q}\,\cos\left(\frac{\theta}{3}\right) 
                            - \frac{p_2}{3}, \nonumber\\
 \lambda_2 &=& -2\,\sqrt{Q}\,\cos\left(\frac{\theta + 4\pi}{3}\right) 
                            - \frac{p_2}{3}, \nonumber\\
 \lambda_3 &=& -2\,\sqrt{Q}\,\cos\left(\frac{\theta + 2\pi}{3}\right) 
                            - \frac{p_2}{3} ,
\end{eqnarray}
where 
\begin{equation}
 \cos\, \theta = P/Q^{3/2},
\end{equation}
with 
\begin{eqnarray*}
 P &=& (p_2/3)^3 - (p_1\, p_2 - 3\, p_0)/6, \\
 Q &=& (p_2/3)^2 - (p_1/3),
\end{eqnarray*}
and 
\begin{eqnarray*}
 p_0 &=& C_{RR}C_{IV}^2 + C_{VV}C_{IR}^2 + C_{II}C_{RV}^2 \\
     && \quad -C_{RR}C_{VV}C_{II} - 2\,C_{IR}C_{IV}C_{RV} \\ 
     &=& - \lambda_1 \lambda_2 \lambda_3,\\
 p_1 &=& C_{RR}C_{VV} - C_{RV}^2 + C_{RR}C_{II} - C_{IR}^2 
         + C_{II}C_{VV} - C_{IV}^2,  \\
     &=& \lambda_1\lambda_2 + \lambda_1 \lambda_3 + \lambda_2\lambda_3\\
 p_2 &=& - (C_{RR} + C_{VV} + C_{II}) = -(\lambda_1 + \lambda_2 + \lambda_3)
\end{eqnarray*}
(e.g. Section 5.6 of Press et al. 2007).

If we write the eigenvector associated with eigenvalue $\lambda_i$ 
as 
\begin{equation}
 {\bm\Lambda}_i = {\bm r} - a_i\, {\bm v} - b_i\, {\bm i},
\end{equation}
where ${\bm r}, {\bm v}$ and ${\bm i}$ are unit vectors in the 
size, velocity dispersion, and surface-brightness directions, 
then 
\begin{eqnarray}
 a_i &=& \frac{C_{RV}/C_{VV}}{1 - \lambda_i/C_{VV}}
               -  b_i\,\frac{C_{IV}/C_{VV}}{1 - \lambda_i/C_{VV}},\\
 b_i &=& 
      \frac{C_{IR}C_{VV}\,(1-\lambda_i/C_{VV}) - C_{IV}C_{RV}}
   {C_{II}C_{VV}(1 - \lambda_i/C_{II})(1 - \lambda_i/C_{VV}) - C_{IV}^2}.
 \label{aibi}
\end{eqnarray}
We are particularly interested in the smallest eigenvalue, since 
the square root of it gives the intrinsic rms scatter orthogonal 
to the Fundamental Plane.  


Suppose this eigenvalue is $\lambda_3$.  
Then the coefficients of the associated eigenvector are given by 
inserting $\lambda_3$ in the expression above.  It is conventional 
to use $(a_{\rm orth},b_{\rm orth})$ to denote $(a_3,b_3)$, so that 
\begin{eqnarray}
 \label{aorth}
 a_{\rm orth} &=& \frac{C_{RV}/C_{VV}}{1 - \lambda_3/C_{VV}}
               -  b_{\rm orth}\,\frac{C_{IV}/C_{VV}}{1 - \lambda_3/C_{VV}}\\
 b_{\rm orth} &=& 
      \frac{C_{IR}C_{VV}\,(1-\lambda_3/C_{VV}) - C_{IV}C_{RV}}
   {C_{II}C_{VV}(1 - \lambda_3/C_{II})(1 - \lambda_3/C_{VV}) - C_{IV}^2},
 \label{borth}
\end{eqnarray}
with intrinsic uncertainty well-approximated by 
$\langle\Delta R^2_{\rm orth}\rangle^{1/2}/(N_{\rm obj}C_{V|I})^{1/2}$ and 
$\langle\Delta R^2_{\rm orth}\rangle^{1/2}/(N_{\rm obj}C_{I|V})^{1/2}$ 
with 
$\langle\Delta R^2_{\rm orth}\rangle
 \equiv (1 + a_{\rm orth}^2 + b_{\rm orth}^2)\,\lambda_3$.  
Measurement errors make these 
$\chi^2_{\rm obs,orth}(O^w_{V|I}/C_{V|I})/C_{V|I}$ 
and 
$\chi^2_{\rm obs,orth}(O^w_{V|I}/C_{I|V})/C_{I|V}$ 
where 
$\chi^2_{\rm obs,orth} = O_{RR} + a_{\rm orth}^2 O_{VV} + b_{\rm orth}^2O_{II} - 2a_{\rm orth}O_{RV} - 2b_{\rm orth}O_{IR} + 2a_{\rm orth}b_{\rm orth}O_{IV}$.

Notice that, in the thin Plane limit, $\lambda_3\to 0$, so 
\begin{eqnarray}
 a_{\rm orth} &\to& \frac{C_{RV}C_{II} - C_{IV}C_{IR}}
                       {C_{II}C_{VV} - C_{IV}^2}\\
 b_{\rm orth} &\to& 
      \frac{C_{IR}C_{VV} - C_{IV}C_{RV}}{C_{II}C_{VV} - C_{IV}^2}.
 \label{thinplane}
\end{eqnarray}
Comparison with equations~(\ref{adirect}--\ref{bdirect}) shows that, 
in this limit, the coefficients of the direct and orthogonal fits are 
the same (as they should be).  

When $C_{IV} \ll 1$ then 
\begin{equation}
 a_{\rm orth} \to \frac{C_{RV}/C_{VV}}{1 - \lambda_3/C_{VV}} 
 \quad {\rm and}\quad
 b_{\rm orth} \to \frac{C_{IR}/C_{II}}{1 - \lambda_3/C_{II}}.
 \label{cIV=0}
\end{equation}
Since $\lambda_3$ is the smallest eigenvalue, it is smaller than 
either $C_{II}$ or $C_{VV}$, so the coefficients of the orthogonal 
fit are {\em guaranteed} to be larger than those of the direct fit; 
in this limit, this means that they are slightly larger than the 
slopes of the simpler pairwise $\langle R|V\rangle$ and 
$\langle R|I\rangle$ relations.  
In practice, $C_{VV}\ll C_{II}$ so this will make 
$a_{\rm orth}>a_{\rm direct}$ but $b_{\rm orth}\approx b_{\rm direct}$.  

These expressions (e.g. equation~\ref{cIV=0}) make it easy to 
understand the effect of restricting the range of $\sigma$ in 
the sample, as is done in Hyde \& Bernardi (2009b).  This will have 
the effect of decreasing $C_{VV}$, making $\lambda_3/C_{VV}\to 1$, 
thus increasing $a_{\rm orth}$, but leaving $b_{\rm orth}$ 
essentially unchanged (see Figure~8 in Hyde \& Bernardi 2009b).

\subsection{The orthogonal fit:  Eigenvectors}\label{evectors}
Although we concentrated on the smallest eigenvalue and its 
eigenvector, the expressions above are also valid for each 
eigenvalue.  Thus, if the largest eigenvalue, $\lambda_1$, 
is much larger than $C_{VV}$, then the associated eigenvector 
${\bm\Lambda}_1$ will have essentially no component in the $V$ 
direction:  $a_1\approx 0$.  
When this is the case, as it is for most datasets 
($\lambda_1$ must be greater than $C_{II}$ and $C_{II}\gg C_{VV}$ 
for most if not all FP datasets), 
then the fact that the three eigenvectors are orthogonal allows us 
to express the coefficients of the other two eigenvectors (those 
in the FP rather than orthogonal to it) as simple combinations of 
$a_{\rm orth}$ and $b_{\rm orth}$.  Namely, 
 ${\bm\Lambda}_3 \cdot {\bm\Lambda}_1 = 0$ sets $b_1=-1/b_{\rm orth}$, 
and then 
 ${\bm\Lambda}_1 \times {\bm\Lambda}_3 = {\bm\Lambda}_2$ sets 
$a_2$ and $b_2$.  This procedure yields equation~(\ref{FPvectors}), 
illustrating that $C_{II}\gg C_{VV}$ plays a key role.  

\subsection{The FP with normalized variables}
One might argue that the real Plane of interest is the one 
obtained by normalizing all observables by their rms values.  
This means that we are interested in the eigenvalues and 
vectors of ${\cal R}$ (c.f. equation~\ref{calR}).
The coefficients of the direct fit become  
 $a_{\rm direct} = (r_{RV} - r_{IR}r_{IV})/(1 - r_{IV}^2) = 0.678$ 
and 
 $b_{\rm direct} = (r_{IR} - r_{RV}r_{IV})/(1 - r_{IV}^2) = -0.700$.
The three eigenvalues are $0.081, 1.123, 1.796$ and the associated 
orthogonal fit coefficients are 
 $(a_{\rm orth},b_{\rm orth}) = (0.75,-0.77)$.  

This Plane is easy to understand if we set $r_{IV}=0$ (this is 
analogous to our setting $C_{IV}/C_{II}\to 0$).  Then 
\begin{equation}
{\cal R} \approx 
 \left(\begin{array}{ccc}
           1 & r_{IR} & 0\\
           r_{IR} & 1 & r_{RV}\\
           0 & r_{RV} & 1\\
          \end{array}\right).
\end{equation}
The associated eigenvalues are $1, 1\pm\sqrt{r_{IR}^2 + r_{RV}^2}$
with eigenvectors 
\begin{eqnarray}
 {\bm\Lambda}_3 &=& {\bm i} - \sqrt{1 + (r_{RV}/r_{IR})^2}\, {\bm r} 
                                  + (r_{RV}/r_{IR})\, {\bm v},\\
 {\bm\Lambda}_2 &=& {\bm i} - (r_{IR}/r_{RV})\, {\bm v},\\
 {\bm\Lambda}_1 &=& {\bm i} + \sqrt{1 + (r_{RV}/r_{IR})^2}\, {\bm r} 
                                  + (r_{RV}/r_{IR})\, {\bm v}.
\end{eqnarray}
Since $r_{IR}\approx -r_{RV}$, this reduces further to 
\begin{eqnarray}
 {\bm\Lambda}_3 &\approx& {\bm i} - \sqrt{2}\, {\bm r} - {\bm v},\\
 {\bm\Lambda}_2 &\approx& {\bm i} + {\bm v},\\
 {\bm\Lambda}_1 &\approx& {\bm i} + \sqrt{2}\, {\bm r} - {\bm v}.
\end{eqnarray}
Notice that the equation for this FP is rather different than 
when the observables were not normalized by their rms values.

\subsection{When one correlation is due to the other two}
The previous section showed the simplifications which are possible if 
one of the pairwise correlations vanishes.  The other case of interest 
is when one of the correlations is entirely due to the other two.  
An example of this is the color-$\sigma$-luminosity relation:  
the color-luminosity correlation is entirely due to that between 
color-$\sigma$ and $\sigma$-luminosity (Bernardi et al. 2005).  
In this case, 
\begin{equation}
{\cal R} \approx 
 \left(\begin{array}{ccc}
           1 & r_{CV} & r_{CV}r_{VL}\\
           r_{CV} & 1 & r_{VL}\\
           r_{CV}r_{VL} & r_{VL} & 1\\
          \end{array}\right),
\end{equation}
where $C$, $V$ and $L$ denote color, $\log(\sigma)$ and log(luminosity),
so $p_0 = -(r_{CV}^2-1)(r_{VL}^2-1)$,
   $p_1 = 3 - r_{CV}^2 - r_{VL}^2 - r_{CV}^2r_{VL}^2$,
and $p_2 = -3$.  This makes 
 $P = -r_{CV}^2r_{VL}^2$ and $Q = (r_{CV}^2 + r_{VL}^2 + r_{CV}^2r_{VL}^2)/3$.
Unfortunately, the expressions for the eigenvalues and vectors which 
result are complicated, and not very intuitive.  

However, they simplify if $r_{CV}=r_{VL}$, in which case 
the three eigenvalues are 
$(r_{CV}^2 + r_{CV}\sqrt{r_{CV}^2 + 8} + 2)/2$, 
$1-r_{CV}^2$, and 
$(r_{CV}^2 - r_{CV}\sqrt{r_{CV}^2 + 8} + 2)/2$,
and the associated eigenvectors are
\begin{eqnarray}
 {\bm\Lambda}_1 &\approx& {\bm l} + {\bm c} -
   \frac{(r_{CV}^2-4) - r_{CV}\sqrt{r_{CV}^2+8}}{\sqrt{r_{CV}^2+8} + 3r_{CV}}
        \,{\bm v}, \\
 {\bm\Lambda}_2 &\approx& {\bm l} - {\bm c} ,\\
 {\bm\Lambda}_3 &\approx& {\bm l} + {\bm c} +  
   \frac{(r_{CV}^2-4) + r_{CV}\sqrt{r_{CV}^2+8}}{\sqrt{r_{CV}^2+8} - 3r_{CV}}
    \,{\bm v}.
\end{eqnarray}
Unfortunately, this is not so useful for interpretting the SDSS data, 
which have $r_{VL}\approx 0.8$ and $r_{CV}\approx 0.5$.  This is one 
example of where direct analysis of the elements of the covariance 
matrix is more interesting, and provides more insight, than analysis 
of its principle components.

\section{Differential evolution effects}\label{evolve}
Our analysis shows that the form of the $z=0$ FP is largely a consequence 
of the fact that the distribution of surface brightness is much larger 
than that in velocity dispersion, and surface brightness and velocity 
are almost uncorrelated.  In passive differential evolution models, 
in which the luminosities of the lower mass galaxies are assumed to 
evolve fastest while sizes and velocity dispersions do not change, 
this is an accident:  surface brightness and velocity dispersion should 
no longer be uncorrelated at $z>0$.  As a result, the coefficients of 
the FP are expected to evolve.  The following simple example illustrates.  

\subsection{Passive luminosity evolution}
Suppose that 
\begin{equation}
 L_z = L_0 (1+z)^{\alpha(M_{\rm dyn})}
\end{equation}
where $M_{\rm dyn}\propto R\sigma^2$ is the same at all redshifts, and 
\begin{equation}
 \alpha(M_{\rm dyn}) = \alpha_* - \beta_* (M_{\rm dyn}-\langle M_{\rm dyn}\rangle).
\end{equation}
The sign has been chosen so that $\beta_*>0$ means massive galaxies 
evolve less rapidly.  
Then, at redshift $z$, the slope of the relation between 
log(dynamical mass) and log(luminosity) will be 
\begin{equation}
 \frac{C_{L_z M_{\rm d}}}{C_{M_{\rm d}M_{\rm d}}} = \frac{C_{L_o M_{\rm d}}}{C_{M_{\rm d}M_{\rm d}}} 
   - \beta_*\,\log(1+z).
\end{equation}
This shows that the slope will decrease at high $z$ if $\beta_*>0$ 
(i.e., if massive galaxies evolve less rapidly).  As a result, the 
slope of log($M_{\rm dyn}/L$) at fixed $M_{\rm dyn}$ (which is one minus 
the number on the right hand side of the expression above) will steepen 
at higher $z$ for positive $\beta_*$.  

\begin{figure}
 \centering
 \includegraphics[width=0.95\hsize]{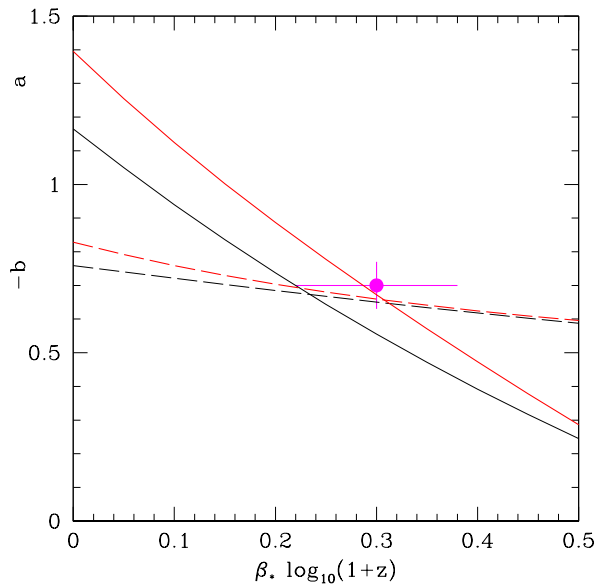}
 \caption{Relation between FP parameters $a$ (solid) and $b$ (dashed) 
          and the change in the slope of the dynamical mass-to-light 
          ratio in a model in which only luminosities evolve, and this 
          evolution depends on dynamical mass at $z=0$:  massive galaxies 
          evolve less rapidly.  Upper (thick) solid and dashed curves 
          are for the orthogonal fit; the thinner solid and dashed curves 
          are for the direct fit.  
          Filled circle and associated error bar shows the measurement 
          of J{\o}rgensen et al. (2006).}
 \label{evolveFP}
\end{figure}

Similarly, although $C_{RV}$, $C_{RM_{\rm d}}$ and $C_{VM_{\rm d}}$ do not 
evolve, correlations which involve luminosity do.  For example, at 
redshift $z$, the correlation between surface brightness and velocity 
dispersion becomes 
\begin{equation}
 C_{I_zV} = C_{L_zV} - 2 C_{RV} = C_{I_0V} - \beta_*\log(1+z) C_{VM_{\rm d}};
\end{equation}
since $C_{VM_{\rm d}}>0$, we expect $C_{I_zV}$ to have the opposite sign 
to $\beta_*$.  In particular, for $\beta_*>0$ we expect $C_{I_zV}<0$, so 
equation~(\ref{adirect}) implies that 
$a_{\rm direct}(z) < a_{\rm direct}(0)$ if $C_{I_zR}/C_{I_zV} > C_{RV}/C_{VV}$.  
Since $C_{RV}\approx C_{VV}$, this means that we would like to know if 
$C_{I_zR} > C_{I_zV}$.  A little algebra, combined with the fact that 
$C_{I_0V}\approx 0$, $C_{VV}<C_{RR}$ and $C_{RV}\approx C_{VV}$ shows that 
$a_{\rm direct}(z) < a_{\rm direct}(0)$ if $\beta_*>0$.  
A similar analysis of equation~(\ref{bdirect}) shows that $b_{\rm direct}$ 
too decreases with $z$ if $\beta_*>0$.  
However, note that for $\beta_*>0$, the distribution of 
surface-brightnesses widens (i.e., $C_{I_zI_z}>C_{I_0I_0}$) meaning 
$C_{I_zI_z}\gg C_{VV}$, so, even though $a_{\rm orth}$ and $b_{\rm orth}$ both 
change, equation~(\ref{FPvectors}) continues to describe the Plane well.  

Notice that, in such models, the evolved values of $(a,b)$ depend on 
the change in the slope of the mass-to-light ratio.  This is shown in 
Figure~1, where we have also shown the expected relation for the 
orthogonal fit coefficients, to illustrate that they behave similarly.  
Though we have not shown it here, the intrinsic scatter also changes 
slightly:  
If we define $\beta_z\equiv \beta_*\log_{10}(1+z)$, then 
$\langle\Delta_{\rm direct}^2\rangle^{1/2}$ decreases from about 0.1 
at $\beta_z=0$ to 0.07 at $\beta_z=0.5$, whereas 
$\lambda_3^{1/2}$ increases from about 0.053 to 0.058. 

For comparison the filled circle shows a measurement of these quantities 
at $z\sim 0.85$, from J{\o}rgensen et al. (2006).  (In fact, we have only 
shown their measurement of the change in slope of 
 $\langle M_{\rm dyn}/L|M_{\rm dyn}\rangle$, $0.3\pm 0.08$, 
versus their measurement of $-b = 0.7\pm 0.07$, which is close to 
what we call $-b_{\rm orth}$.  
They also report $a = 0.6\pm 0.22$, which would be displaced slightly 
downwards on our plot, and have substantially larger uncertainties, than 
the single point we have shown.)  Note that their measurement of the 
change in slope implies $\beta_*\approx 0.3/\log_{10}(1.9)\approx 1.07$.  
They also report little change in the thickness of the plane, which 
is consistent with the numbers given above.  If this is indeed the 
right picture, then the luminosity function at $z$ should be narrower 
by a factor of 
$C_{L_zL_z}/C_{L_0L_0} = 1 - 2\beta_z C_{L_oM_d}/C_{L_0L_0} 
                       + \beta_z^2 C_{M_dM_d}/C_{L_0L_0} 
                   \approx 0.5$.

Before we move on, it is worth remarking on the fact that differential 
luminosity evolution changes $a$ more than $b$.  Naively, this is 
surprising, since $a_{\rm direct}\approx C_{RV}/C_{VV}$ at $z=0$, so one 
might have thought it would not be changed at all if neither $R$ nor $V$ 
change.  Moreover, one might have expected $b$ to change, perhaps strongly, 
because the luminosity evolution would change both $C_{IR}$ and $C_{II}$.  
To see why $b$ changes only weakly, note that $\beta_*>0$ means that 
the distribution of $L$ was narrower at high $z$.  In the limit in 
which all objects have the same luminosity $C_{IR}/C_{II} = -1/2$; 
thus, differential evolution cannot force $|b|$ below $1/2$.  
Since $|b|=0.8$ at $z=0$, and it cannot become smaller than 1/2, the 
evolution in $b$ is weak.  
Thus, our analysis shows that $a$ is more strongly affected than $b$ 
because luminosity evolution makes $C_{IV}\ne 0$ at higher $z$, and 
because differential evolution makes the distribution of $L$ narrower 
in the past.  

\subsection{Selection effects and structural evolution}
While consistent with the measurements, pure (differential) luminosity 
evolution is not required by them.  
For example, the expected form of this evolution implies a narrower 
distribution of $L$ at high redshift.  Since a magnitude limited 
selection effect would also produce a narrower distribution of $L$,  
one must first be sure that this is not producing the observed 
changes in $a$ and $b$.  In particular, Figure~7 in Hyde \& Bernardi (2009b) 
shows that removing faint galaxies from the $z=0$ sample decreases $a$ 
and $|b|$.  Since this is qualitatively the same as the change in the 
FP coefficients between $z=0$ and $z=0.9$, statements about differential 
evolution should only be believed if accompanied by measurements of a 
change in the slope of the size-$L$ and $\sigma-L$ relations -- the 
FP itself is a very bad diagnostic.  

Moreover, the analysis above assumes that only the luminosities evolve.  
However, there is much recent discussion of the fact that, at fixed 
stellar mass, galaxies appear to be more than three times smaller at 
$z\sim 2$ than at $z\sim 0$ 
(e.g. Trujillo et al. 2006; Cimatti et al. 2008; Van Dokkum et al. 2008) 
although the evidence is not uncontested (e.g. Mancini et al. 2010; 
Sarocco et al. 2010).  
Indeed, Saglia et al. (2010) interpret their measurements of the 
evolution of the Fundamental Plane entirely in terms of structural 
evolution, rather than differential evolution of luminosity!
  
At fixed $M_{\rm dyn}$, they find that the sizes are slightly smaller 
and velocity dispersions slightly larger at $z\sim 0.8$ than at 
$z\sim 0$.  While the redshift dependance they report is in quantitative 
agreement with that derived by Bernardi (2009) from a substantially larger 
dataset restricted to a narrower redshift range ($z<0.3$), we must 
again worry about selection effects on these estimates of structural 
evolution.  
For example, suppose that the evolution was purely in the luminosities, 
and it was not differential, but the high-$z$ measurements only see the 
largest $L$.  Then because both $R$ and $M_{\rm dyn}$ correlate with $L$, 
the $R-M_{\rm dyn}$ relation will be biased by this selection on $L$ 
(even though $L$ does not enter explicitly in the 
$\langle R|M_{\rm dyn}\rangle$ relation).  
In addition, relating the high-$z$ measurements to those at $z=0$ 
requires a better understanding of the systematic differences in 
band-passes, of how the velocity dispersion measurement at high-$z$ 
relates to the one at $z=0$ (e.g., effective aperture effects), and 
of whether or not the high-$z$ population really is made up of the 
progenitors of the $z=0$ population.  Exploring this further 
(e.g. How should one account for the fact that the youngest members of 
the $z=0$ population simply did not exist at $z\sim 1$?
What role do mergers play?), in the context of differential evolution 
models, is the subject of work in progress.  


\section{Discussion}
We started from a general expression for the conditional distribution 
of $n$ correlated variables when $N-n$ other variables are known 
(equations~\ref{condMean} and~\ref{condCovariance}), and specialized to 
the case $N=3$.  This provided analytic expressions which describe the 
Fundamental Plane associated with three correlated variables.  
Our expressions allow one to see why the coefficients of the direct, 
inverse and orthogonal fits differ (equations~\ref{adirect}--\ref{bdirect}, 
\ref{ainverse}--\ref{binverse}, \ref{aorth}--\ref{borth}, and 
Table~\ref{tab:FPfits}); 
how to estimate the uncertainties on these coefficients; 
why the three eigenvectors which describe the FP have the form they 
do (equation~\ref{FPvectors} and Section~\ref{evectors});
and to see how and why the Fundamental Plane in a magnitude limited 
survey will, in general, differ from that in a complete sample 
(Appendix).  

If one views all pairwise correlations as having a component that 
is due to the individual correlations between each observable and 
luminosity, and another component which is not, 
then our analysis shows that only the part which is not due to the 
correlations with luminosity remains unaffected by the magnitude 
limited selection:  the other part is biased (e.g., equation~\ref{invariant}).  
Our analysis also shows how to remove this bias, as well as 
account for measurement errors.  By providing analytic expressions 
for all quantities of interest, our results remove the need for
numerical nonlinear minimization methods for obtaining the best-fit 
coefficients.  
These results were used by Hyde \& Bernardi (2009b) in their 
analysis of the SDSS Fundamental Plane.  

Many properties of the Fundamental Plane at $z=0$ can be understood 
as arising from the fact that surface brightness and velocity dispersion 
are uncorrelated.  This raises the question of whether or not this 
lack of correlation encodes something fundamental about the physics 
of galaxy formation.  
Recent work suggests that the coefficients of the Fundamental Plane 
at $z=0.8$ are significantly different from those at $z=0$ 
\cite{dSALJ,hiz}.  
We showed that, in models where massive galaxies evolve less rapidly 
than low mass galaxies, but there are no changes to the size or 
velocity dispersions, there is a one-to-one relation between the 
changes to $(a,b)$ and the correlation between luminosity and mass 
(Figure~\ref{evolveFP}).  (We also showed that, even though $(a,b)$ 
change, the relationship between the eigenvectors of the Plane 
(equation~\ref{FPvectors}) does not.)  
This relation, which is in reasonable agreement with the measurements, 
also predicts that $C_{IV}\ne 0$ at higher $z$.  I.e., in this model, 
$C_{IV}=0$ at $z=0$ is just a coincidence.  

While consistent with the FP measurements, pure (differential) luminosity 
evolution is not required by them.  E.g., a selection effect on 
luminosity will produce qualitatively similar changes to $a$ and $b$, 
making the FP a very bad diagnostic of this sort of evolution; 
the size-$L$ and $\sigma-L$ relations are much better.  
Moreover, other scaling relations suggest there has been substantial 
structural evolution since $z\sim 1$.  Again, selection effects 
complicate the relationship between the observed changes to $a$ and 
$b$, and the structural evolution parameters.  Accounting for these 
is the subject of work in progress, but we note that if $C_{IV}$ 
remains small even at high $z$, then this will provide a simple way to 
constrain models of the structural changes that complement differential 
luminosity evolution.

\section*{Acknowledgements}
We thank the organizers of the meeting held in Ensenada, Mexico 
in March 2008 for inviting us to attend, which prompted us to 
complete this work, the organizers of the Cosmic Comotion workshop 
on Stadbroke Island in September 2010 which prompted us to submit, 
and P. Schechter for suggesting the title during a visit to the IAS 
many years ago.  We would also like to thank the referee for a very 
helpful report, and for identifying a number of typos in the original 
version of this paper.
This work was supported in part by NASA grant ADP/NNX09AD02G to 
MB, and by NSF-AST 0908241 to RKS.

\appendix

\section{Biases from the flux-limited selection effect}
The discussion in the main text can be worked through for the 
case of an apparent magnitude-limited survey in which one does 
not weight objects by (the inverse of) $V_{\rm max}(L)$.  In essence, 
all one must do is determine the change to the elements of the 
covariance matrix if all objects have the same weight.  
Although the main text worked with luminosity in solar units, 
rather than absolute magnitudes, the analysis in this Appendix 
uses magnitudes.  We use $M\propto -2.5\log_{10}(L)$ for absolute 
magnitude -- it should not be confused with $M_{\rm d}$ in the 
main text, which we used for dynamical mass -- and so now surface 
brightness is $I\propto M + 5R$.  

\subsection{Quantifying the bias}
If we use $\bar X$ and $\bar C_{XY}$ to denote the means and 
(error-corrected) covariances in the observed sample (i.e. 
equations~\ref{meanX} and~\ref{Cxy} with $w_i=1$ for all $i$), 
then the fact that $\bar C_{XY}\ne C_{XY}$ for all pairs $XY$ means 
that the coefficients of the Fundamental Plane are sensitive to 
selection effects, so care must be taken when estimating its shape.  
When there is no curvature in the underlying pairwise scaling 
relations, then this is straightforward, as we show below.  
In essence, all that is really required is an estimate of how the 
mean and the width of the observed luminosity distribution is affected 
by the magnitude-limited selection.  

For example, the differences between the selection-biased and intrinsic 
mean values are given by 
\begin{eqnarray}
 \bar R - \langle R\rangle 
    &=& \frac{C_{RM}}{C_{MM}}\,\Bigl(\bar M - \langle M\rangle\Bigr),\qquad
 \bar I = \bar M + 5\,\bar R, \nonumber\\
 \bar V - \langle V\rangle 
   &=& \frac{C_{VM}}{C_{MM}}\,\Bigl(\bar M -  \langle M\rangle\Bigr),
\end{eqnarray}
where $\langle M\rangle$ etc. denote the true mean values 
(i.e., those in which the selection effect has been accounted-for).
Similarly, the selection-biased covariances are 
\begin{eqnarray}
 \bar C_{RM} &=& \frac{C_{RM}}{C_{MM}}\,\bar C_{MM}, \qquad 
 \bar C_{VM}  =  \frac{C_{VM}}{C_{MM}}\,\bar C_{MM}, \nonumber\\
 \bar C_{RR} &=& C_{RR} + \frac{C_{RM}^2}{C_{MM}^2}\, 
                    \left(\bar C_{MM} - C_{MM}\right),\nonumber\\
 \bar C_{VV} &=& C_{VV} + \frac{C_{VM}^2}{C_{MM}^2}\, 
                    \left(\bar C_{MM} - C_{MM}\right),\nonumber\\
 \bar C_{RV} &=& C_{RV} + \frac{C_{RM}C_{VM}}{C_{MM}^2} 
                    \left(\bar C_{MM} - C_{MM}\right),
\end{eqnarray}
from which one can compute 
\begin{eqnarray}
 \bar C_{IM} &=& \bar C_{MM} + 5\,\bar C_{RM}, \nonumber\\ 
 \bar C_{IR} &=&  \bar C_{RM} + 5\,\bar C_{RR}, \qquad
 \bar C_{IV}  =   \bar C_{VM} + 5\,\bar C_{RV}, \nonumber\\ 
 \bar C_{II} &=& \bar C_{MM} + 10\,\bar C_{RM} + 25\,\bar C_{RR}.
\end{eqnarray}
This shows that scaling relations at fixed $M$ are not affected by 
the selection effect:  $\bar C_{RM}/\bar C_{MM} = C_{RM}/C_{MM}$ etc.  
For the other relations, the differences from when $V_{\rm max}^{-1}$ 
weighting is used depend on how different $\bar C_{MM}$, the variance 
in the observed luminosity distribution, is from the intrinsic 
variance, $C_{MM}$.  This difference will differ from one sample to 
another:  we will quantify it for the SDSS sample shortly.  

\subsection{Correcting the bias}
These expressions can be rearranged to express the correct intrinsic 
correlations in terms of the selection-biased ones:  
\begin{eqnarray}
 C_{RM} &=& \frac{\bar C_{RM}}{\bar C_{MM}}\,C_{MM}, \qquad 
 C_{VM}  =  \frac{\bar C_{VM}}{\bar C_{MM}}\,C_{MM}, \nonumber\\
 C_{RR} &=& \bar C_{RR} - \frac{\bar C_{RM}^2}{\bar C_{MM}^2}\, 
                    \left(\bar C_{MM} - C_{MM}\right),\nonumber\\
 C_{VV} &=& \bar C_{VV} - \frac{\bar C_{VM}^2}{\bar C_{MM}^2}\, 
                    \left(\bar C_{MM} - C_{MM}\right),\nonumber\\
 C_{RV} &=& \bar C_{RV} - \frac{\bar C_{RM} \bar C_{VM}}{\bar C_{MM}^2} 
                    \left(\bar C_{MM} - C_{MM}\right).
\end{eqnarray}
The intrinsic correlations with $I$ can then be got from 
\begin{eqnarray}
 C_{IM} &=& C_{MM} + 5\,C_{RM}, \nonumber\\
 C_{IR} &=& C_{RM} + 5\,C_{RR}, \qquad
 C_{IV}  =  C_{VM} + 5\,C_{RV}, \nonumber\\
 C_{II} &=& C_{MM} + 10\,C_{RM} + 25\,C_{RR},
\end{eqnarray}
with mean values 
\begin{eqnarray}
 \langle R\rangle 
    &=& \bar R - \frac{\bar C_{RM}}{\bar C_{MM}}\,
        \Bigl(\bar M - \langle M\rangle\Bigr),\qquad
 \langle I\rangle = \langle M\rangle + 5\,\langle R\rangle, \nonumber\\
 \langle V\rangle 
   &=& \bar V - \frac{\bar C_{VM}}{\bar C_{MM}}\,
       \Bigl(\bar M -  \langle M\rangle\Bigr).
\end{eqnarray}
Note that the quantity which is the same in the full and magnitude 
limited samples is 
\begin{eqnarray}
 \bar C_{RV} - \frac{\bar C_{RM} \bar C_{VM}}{\bar C_{MM}}
  &=& C_{RV} - \frac{C_{RM}C_{VM}}{C_{MM}} \nonumber\\
  &=& C_{RV}\,\frac{r_{RV} - r_{RM}r_{VM}}{r_{RV}}.
 \label{invariant}
\end{eqnarray}
This makes intuitive sense, because the expression above is the 
part of the correlation between $R$ and $V$ which is not due to 
the individual correlations between $R$ and $M$, and $V$ and $M$.  
This part, i.e., the part which does not correlate with $M$, remains 
unchanged by the magnitude limited selection.  Similar relations hold 
for $C_{RR}$, $C_{VV}$, etc.

The analysis above shows that, to account for the selection bias, 
all one needs is an estimate of the difference between the unweighted 
and weighted mean and variance of the absolute magnitude distribution 
(i.e. of the bias in the luminosity function).  
In the SDSS dataset of Hyde \& Bernardi (2009b), 
\begin{eqnarray}
 \bar M &=& -21.94, \qquad \qquad \bar C_{MM} = 0.65, \nonumber\\
 \langle M\rangle &=& -20.99,\qquad {\rm and}\quad C_{MM} = 0.76,
\end{eqnarray}
So, e.g., $\bar C_{RR} < C_{RR}$ and $\bar C_{VV} < C_{VV}$.  
This illustrates a trivial but important point:  the width of 
the luminosity (and other) distributions in a magnitude limited 
catalog -- i.e., before correcting for the selection effect -- may be 
{\em narrower} than in the intrinsic distribution. 

The expressions above also show that the magnitude limited catalog 
can exhibit correlations between variables even when there is no true 
intrinsic correlation.  E.g, 
\begin{equation}
 \bar C_{IV} 
   = C_{IV} + (\bar C_{MM} - C_{MM}) \frac{C_{VM}}{C_{MM}}\,
              \left(1 + 5\frac{C_{RM}}{C_{MM}}\right);
\end{equation}
thus, $\bar C_{IV}\ne 0$ even if $C_{IV}=0$.  
For similar reasons, absence of a correlation in the magnitude limited 
catalog does not imply vanishing correlation in the full sample.  

We have verified that the expressions above agree with measurements 
of the bias in mock catalogs in which there is no curvature in the 
underlying scaling relations.  In practice, however, there is weak 
curvature in most scaling relations (e.g., Hyde \& Bernardi 2009a; 
Bernardi et al. 2011), 
and this renders the expressions above only approximate.  For example, 
Hyde \& Bernardi (2009b) report that
 $\bar R   = 0.62$, $\bar V = 2.3$ and $\bar\mu = 19.71$, 
 $\bar C_{II} = 0.2660/2.5^2$, $\bar C_{RR} = 0.0488$, $\bar C_{VV} = 0.0127$,  
 $\bar C_{IR} = -0.0820/2.5$, $\bar C_{IV} = -0.0036/2.5$ and 
 $\bar C_{RV} = 0.0159$.  
These are not quite the same as one expects from the expressions above, 
although the differences can be understood in terms of how the underlying 
scaling relations curve.  Nevertheless, our analysis does serve to 
illustrate which relations are expected to be insensitive to selection 
effects arising from a magnitude limit, and which are not.  

\subsection{(In)sensitivity to the bias}
For example, it is sometimes stated that the parameters of the inverse 
fit (equation~\ref{abinv}) and the fit in which $I$ is the dependent 
variable (equation~\ref{abI}) are not affected by the selection effect.  
The analysis above shows that this is, in general, not correct.
However, if we ignore the selection effect then 
 $(\bar a_{\rm inv},\bar b_{\rm inv}) = (1.59,-0.716)$; 
Table~\ref{tab:FPfits} shows that the correct values are $(1.606,-0.792)$, 
suggesting that $a_{\rm inv}$ at least is not very biased, at least 
in the SDSS dataset.  In addition, 
 $I - \bar I = (1.23\pm 0.04)\,(V - \bar V) - (1.07\pm 0.02)\,(R - \bar R)$ 
whereas the parameters from Table~\ref{tab:FPfits} show that 
 $I - \langle I\rangle = 
  1.18\,(V - \langle V\rangle) - 0.97\,(R - \langle R\rangle)$.  
For comparison, Graves \& Faber (2010) report $(1.16, -1.21)$, 
for a slightly different early-type galaxy sample.

In all cases, $a$ is not strongly affected by the magnitude limit.  
To see why, note that 
\begin{eqnarray}
 \frac{\bar C_{RV}}{\bar C_{VV}} 
 &=& \frac{C_{RV}}{C_{VV}}\frac{1 + (C_{RM}C_{VM}/C_{MM}C_{RV})(\Delta C_{MM}/C_{MM})}
                               {1 + (C_{VM}^2/C_{MM}C_{VV}) (\Delta C_{MM}/C_{MM})}
 \nonumber\\
 &\to& \frac{C_{RV}}{C_{VV}} \frac{1 - 5C_{RM}/C_{MM}\,(\Delta C_{MM}/C_{MM})}
                                 {1 + r_{RV}^2 (\Delta C_{MM}/C_{MM})}
\end{eqnarray}
where we have defined $\Delta C_{MM}\equiv \bar C_{MM} - C_{MM}$, 
and the final expression holds in the limit $C_{IV}\to 0$, in which 
case $C_{VM}\to -5C_{RV}$.  Now, $C_{RM}/C_{MM}$ is the slope of the 
size-absolute magnitude relation:  in the SDSS, this is about $-0.24$.  
Similarly, $\Delta C_{MM}/C_{MM}\approx -1/7$ and $r_{VM}\approx 0.8$, 
so the net effect is to have $\bar C_{RV}/\bar C_{VV}$ within about ten 
percent of $C_{RV}/C_{VV}$, making $\bar a_{\rm direct}\approx a_{\rm direct}$ 
also to within about ten percent.  
Since $a_I=a_{\rm direct}$ when $C_{IV}=0$, we expect 
$\bar a_I\approx a_I$, to within ten percent.  
A similar analysis of $a_{\rm inv}$ shows why it too is not strongly 
affected by the magnitude limit.  

\subsection{Biased estimates of the evolution of the zero-point}
Finally, it is worth emphasizing that, although we have focussed on 
the slopes of the correlations, the fact that the mean values in the 
magnitude-limited sample differ from the correct values 
($\bar V\ne \langle V\rangle$ etc.) means that the zero-points of the 
relations can be affected even if the slopes are not.  Since the 
zero-point of the Fundamental Plane is often used as a basis for 
estimating evolution, this estimate must be made carefully in 
magnitude limited samples.  Bernardi et al. (2003) show that this 
effect does indeed produce a significant offset in the SDSS.  
Because we have shown how the mean values and slopes are affected by 
the magnitude limit, our analysis provides a straightforward way to 
correct for this effect.  

Perhaps as importantly, our analysis shows that, just because a 
scaling relation is independent of the magnitude limited selection 
effect at one redshift, there is no guarantee that it will remain 
insensitive at other $z$.  As a specific example, consider the case 
of differential luminosity evolution.  In the main text, we showed 
that if $C_{IV}\approx 0$ at $z=0$, then $C_{IV}\ne 0$ at $z>0$ is 
guaranteed.  However, $C_{IV}=0$ played a crucial role in the previous 
subsection, when we showed why $a$ was insensitive to the magnitude 
limited selection, so at $z>0$, this is no longer guaranteed.

\label{lastpage}

\end{document}